\documentclass[journal=jacsat,manuscript=article]{achemso}

\usepackage{chemformula} % Formula subscripts using \ch{}
\usepackage[T1]{fontenc} % Use modern font encodings

\usepackage{siunitx}

\usepackage{csquotes}
\MakeOuterQuote{"}
\usepackage{bm}
\usepackage{braket}
\renewcommand{\v}[1]{\bm{\mathrm{#1}}}

\def\sss{\scriptscriptstyle\rm}
\def\br{{\bf r}}
\def\bk{{\bf k}}
\def\xc{_{\sss XC}}
\def\ext{_{\rm ext}}
\def\s{_{\sss S}}
\def\H{_{\sss H}}

\author{Sangeeta Sharma}
\affiliation[BigPharma]
{Max-Born-Institute for Non-Linear optics, Max-Born Strasse 2A, 12489 Berlin, Germany}
\author{Deepika Gill}
\affiliation[BigPharma]
{Max-Born-Institute for Non-Linear optics, Max-Born Strasse 2A, 12489 Berlin, Germany}
\author{Jyoti Krishna}
\affiliation[BigPharma]
{Max-Born-Institute for Non-Linear optics, Max-Born Strasse 2A, 12489 Berlin, Germany}
\author{Eddie Harris-Lee}
\affiliation[BigPharma]
{Max-Planck-Institut fur Mikrostrukturphysik Weinberg 2, D-06120 Halle, Germany}
\author{John Kay Dewhurst}
\affiliation[BigPharma]
{Max-Planck-Institut fur Mikrostrukturphysik Weinberg 2, D-06120 Halle, Germany}
\author{Sam Shallcross}
\email{shallcross@mbi-berlin.de}
\affiliation[BigPharma]
{Max-Born-Institute for Non-Linear optics, Max-Born Strasse 2A, 12489 Berlin, Germany}

\title[An \textsf{achemso} demo]
{Giant moment increase by ultrafast laser light}

\abbreviations{IR,NMR,UV}

\begin{document}

\begin{tocentry}

\includegraphics[width=0.86\textwidth]{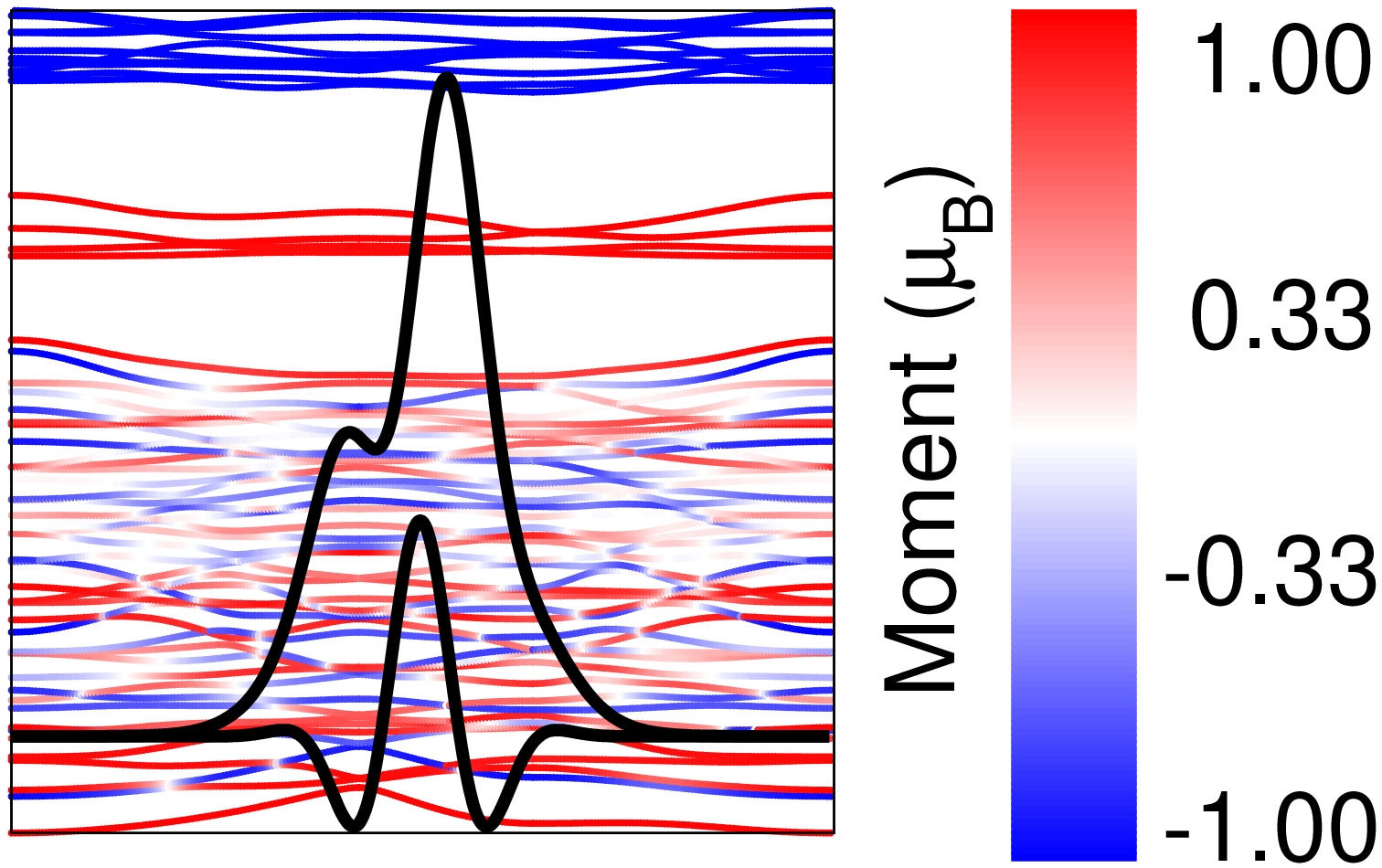}

\end{tocentry}

\begin{abstract}
It is now well established that a few femtosecond laser pulse will induce an ultrafast loss of moment in a magnetic material. Here we show that the opposite effect can also occur: an ultrafast increase in moment. Employing both tight-binding and state-of-the-art time dependent density functional theory we find that laser light tuned to the majority spin conduction band in the 2d magnets CrI$_3$ and CrSBr generates an ultrafast giant moment increase, of up to 33\% in the case of CrI$_3$ (2~$\mu_B$). Underpinning this is spin-orbit induced valence band spin texture that, in combination with a strong field  light pulse, facilitates an optical spin flip transition involving both intra- and inter-band excitation. Our findings, that establish a general mechanism by which ultrafast light pulses may enhance as well as decrease the magnetic moment, point towards rich possibilities for light control over magnetic matter at femtosecond times.
\end{abstract}

\section{Introduction}

It is almost paradigmatic that the moments of transition metal magnets and their compounds will decrease upon irradiation by an ultrafast pulse of intense laser light
\cite{beaurepaire_ultrafast_1996,
malinowski_control_2008,
krieger_laser-induced_2015,
yamamoto_ultrafast_2019,
chekhov_ultrafast_2021}.
While longer (picosecond to nanosecond) time scales can unveil a contrary picture, for example a light induced increase in Curie temperature\cite{disa_photo-induced_2023}, such effects are driven by lightwave coupling to the lattice degrees of freedom and, at the attosecond to few femtosecond times dominated by purely electronic interactions, such physics is precluded. Control over the strength of magnetic order by light thus appears -- for the ultrafast few femtosecond in which interesting coherent quantum effects may occur -- to stand comparatively impoverished.

Here we demonstrate that, surprisingly, for the two dimensional (2d) magnets CrI$_3$\cite{huang_layer-dependent_2017,huang_electrical_2018,
soriano_magnetic_2020} and CrBrS\cite{liu_three-stage_2022,
liu_three-stage_2022-1,
lopez-paz_dynamic_2022,
ziebel_crsbr_2024} this is not true: a giant and ultrafast increase in magnetic moment can be induced by a laser pulse, with an enhancement of up to $2\mu_B$ found in CrI$_3$. Underpinning this effect is the fact that strong spin-orbit coupling imbues the valence band manifolds of these materials with a "spin texture", in which the direction of the moment can evolve from up to down within a single band\cite{soriano_magnetic_2020}. A light pulse generating {\it intra}-band evolution of crystal momentum can thus induce {\it intra}-band spin rotation within the valence band. This forms the basis of an optical excitation -- which as a convenient shorthand we denote an "intraflip" -- combining an intra-band rotation and inter-band excitation that flips the spin while, just as in a direct optical transition, preserving the crystal momentum.

We show that such "intraflips" can drive a giant moment increase at time scales from the single cycle limit to much longer multi-cycle pulses on $\sim$100~fs time scales, with both linearly and circularly polarized light generating moment increase; the effect should therefore be observable employing present day "standard" ultrafast laser pulses. {\it Ab-initio} calculation of the transient XMCD from the Cr L$_{2,3}$ edge reveals clear signatures of this moment increase, allowing experimental probing of our predictions. Requiring only (i) spin-orbit induced spin texture in the valence band and (ii) a pure majority spin conduction band the effect we propose is general, and will apply to any material that fulfils these conditions. Our work thus both highlights the control possibilities of light wave coupling to a non-trivial spin texture, as well as revealing an unexpected richness of magnetic control possible in the ultrafast regime of laser induced spin dynamics.

\section{Ultrafast increase in magnetic moment}

\begin{figure}[t!]
\begin{center}
\includegraphics[width=1\textwidth]{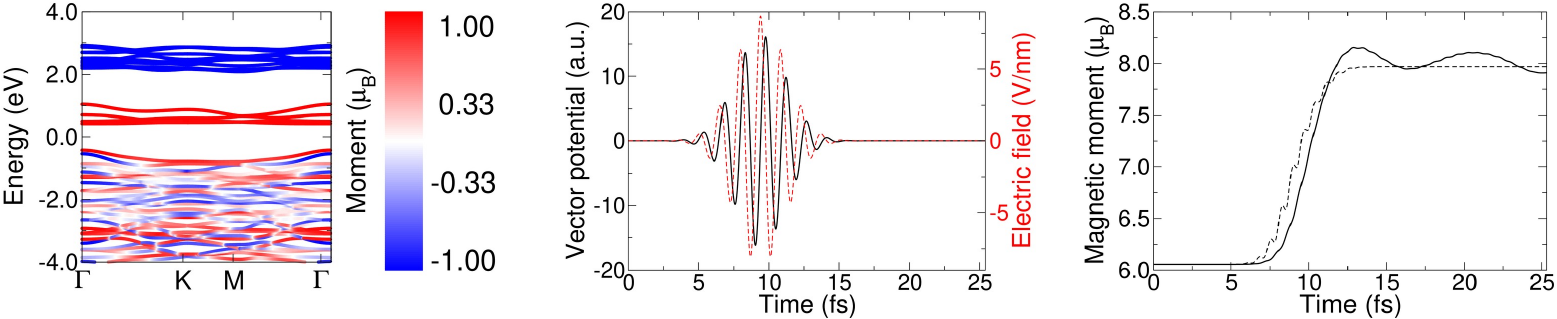}
\caption{\small {\it Ultrafast light induced increase of magnetic moment in CrI$_3$}. (a) The band structure of CrI$_3$ with colour denoting the z-component of spin moment (the other components are zero). Irradiation by a 2.8~eV linearly polarized pulse, vector potential shown in panel (b), generates an increase in magnetic moment from the ground state value of 6~$\mu_B$ to 8~$\mu_B$, panel (c). Here the full line denotes the expectation value of the spin operator (i.e. the full magnetic moment), with the broken line the intra-band contribution to this moment.
}
\label{fig1}
\end{center}
\end{figure}

We first consider the two dimensional semi-conductor CrI$_3$\cite{huang_layer-dependent_2017,huang_electrical_2018}, whose ferromagnetic order (Curie temperature 45~K) has revealed a rich physics of low dimensional magnetism with remarkable light-matter interaction\cite{seyler_ligand-field_2018,
luo_perfect_2021,cheng_light_2021,
song_spin_nodate} and a rich coupling between magnetic, structural, and electronic order, both in a moir\'e geometry\cite{ghader_whirling_2022,
xie_twist_2022,song_direct_2021,
luo_perfect_2021-1,sivadas_stacking-dependent_2018} and in single layer systems\cite{farooq_switchable_2019,
jiang_spin_2018}. While this low Curie temperature will preclude long time measurement of light induced changes in magnetism, as orientational disorder will set in at picosecond times, the robust local moment permits observation of ultrafast spin dynamics via XMCD or MOKE transient spectroscopy. Subsequently, we will demonstrate that similar physics is found in the recently discovered 2d magnet CrBrS\cite{liu_three-stage_2022,
liu_three-stage_2022-1,
lopez-paz_dynamic_2022,
ziebel_crsbr_2024}, that possesses a significantly higher Curie temperature of 132~K. 

To explore the ultrafast spin dynamics of these two materials we employ a Wannierized tight-binding scheme, with key results verified by {\it ab-initio} full potential time-dependent density functional theory. Computational details of both these approaches are presented in the Supplemental document.

The band structure of CrI$_3$ is shown in Fig.~\ref{fig1}(a), with the band colour indicating the spin moment (the Fermi energy is set to zero). A comparison of the tight-binding band structure with that generated by the full potential Elk code can be found in the Supplemental. Two distinctive features can be observed: (i) the conduction bands are either pure spin up or pure spin down while (ii) the valence bands exhibit a more complex spin structure, with continuous evolution of spin up to spin down character within the same band.

Excitation by a 2.8~eV linearly polarized laser pulse, vector potential shown in panel (b), yields, in dramatic contrast to the "expected" behaviour of demagnetization, an increase in moment from 6.06~$\mu_B$ to $\sim $8~$\mu_B$, panel (c). It should be stressed that direct optical transitions cannot generate a change of moment, as such transitions are spin preserving. A quite different mechanism of charge excitation must therefore underpin the ultrafast giant moment increase seen here.

\begin{figure}[t!]
\includegraphics[width=0.65\textwidth]{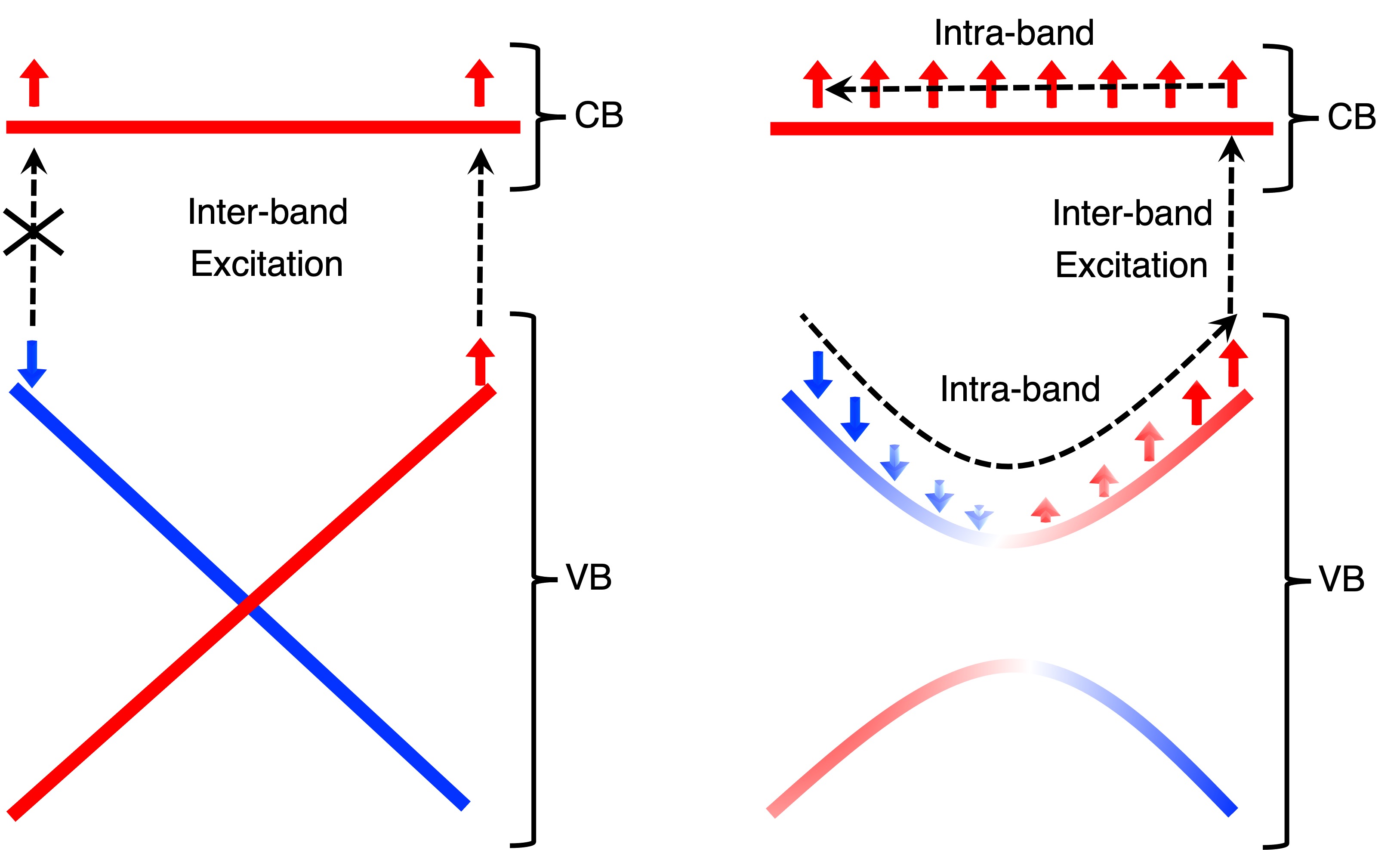}
\caption{\small {\it Intraflip mechanism underpinning light induced magnetic moment increase}. In the absence of spin-orbit coupling optical excitation from spin down to spin up bands is forbidden, panel (a). However, if strong spin-orbit coupling imparts a spin texture to the spin hybridized valence bands (VB), panel (b), then an excitation process in which a spin down state is excited to spin up state is made possible, indicated by the broken line. In this three stage mechanism (i) light induced intra-band excitation evolves a spin state up to a spin down state {\it within} the valence band, followed by (ii) inter-band excitation to the conduction band (CB), allowed as both states are now spin up, with (iii) intra-band evolution then returning this state to its initial crystal momentum while retaining the spin up character as the CB is pure spin up. The overall result is thus a spin flip excitation that preserves the crystal momentum.
}
\label{fig2}
\end{figure}

This is provided for by the fact that strong light pulses generate intra-band evolution of crystal momentum, and not only direct transition from valence to conduction. This excitation mechanism is illustrated schematically in Fig.~\ref{fig2}, where we consider the simplest possible representation of the CrI$_3$ band structure: spin hybridized valence bands and a pure spin up conduction band. In the absence of spin-orbit coupling opposite spin bands are orthogonal, and excitation from the spin down valence band to the spin up conduction band forbidden, panel (a). Strong spin orbit coupling dramatically changes this: valence bands now evolve spin character continuously from up to down, allowing an optical excitation that can flip the spin, indicated in Fig.~\ref{fig2}(b) by the broken line. This consists of a sequence of three excitation steps: (i) intra-band evolution of a spin down state to a spin up state within the valence band; (ii) inter-band excitation to the conduction band; and finally (iii) intra-band motion returning the state to its initial momentum while preserving the spin up state. The net result is an optical transition that, at a given crystal momentum $\v k$, excites a spin down valence state to a spin up conduction band state.

To date ultrafast (i.e. tens of femtoseconds) spin dynamics has been understood in terms of charge excitation driven by direct optical excitation, followed by spin-orbit induced spin flips. This "two step" scheme underpins present day understanding of ultrafast demagnetization\cite{krieger_laser-induced_2015,
krieger_ultrafast_2017,
dewhurst_laser-induced_2018}.
The picture that emerges here is quite different: "intraflip" excitations, via non-perturbative intra-band evolution of momentum, exploit the spin-orbit induced spin structure of the ground state to generate a light induced increase in moment.

\section{Moment control}

\begin{figure}[t!]
\includegraphics[width=1.0\textwidth]{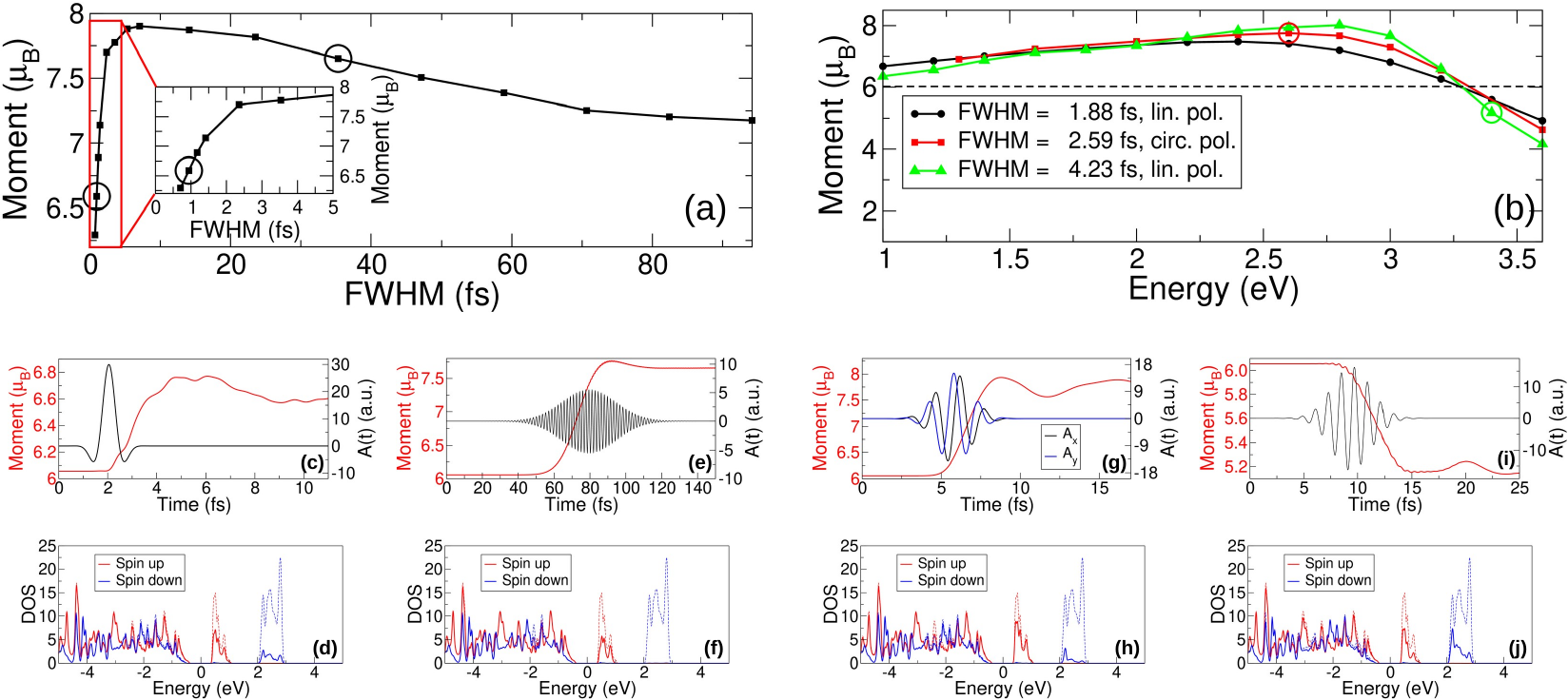}
\caption{\small {\it Ultrafast moment control in CrI$_3$}. (a) Variation of pulse duration while holding the fluence (5.84~mJ/cm$^2$) and central frequency (2.4~eV) fixed reveals that light induced increase in moment occurs from sub-cycle attosecond times to long time multi-cycle pulse, with the inset figure a zoom of the few femtosecond regime. (b) Variation of frequency unveils regimes of both increase and decrease of moment: below $\sim3.2$~eV the pulse generates an increase in moment, while for higher frequencies a reduction in moment is observed. Panels (c-d) and (g-j) present details of representative cases from panel (a) and (b) respectively, as indicated by the open circles in panels (a,b). For each case is presented the vector potential, time dependent moment, and density of states after the pulse.
}
\label{fig3}
\end{figure}

Having established a light induced ultrafast increase in moment we now explore (i) the temporal limits of this phenomena and (ii) how pulse parameters can be used to exert control over the moment dynamics. To this end we first vary the laser pulse duration while holding the frequency fixed at 2.4~eV. We further fix the fluence to a value of 5.84~mJ/cm$^2$, thereby enforcing a "trade off" between pulse amplitude and duration;  longer pulses, to preserve fluence, must have reduced amplitude. As it is the pulse vector potential amplitude determines the magnitude of the light induced intra-band momentum evolution, the "intraflip" mechanism outlined above would suggest that trading pulse amplitude for duration in this way would reduce the efficiency of the laser induced moment increase. This can be seen in Fig.~\ref{fig3}(a), in which a reduction in the moment enhancement occurs on increasing pulse duration: a moment enhancement of 7.90$\mu_B$ at 8~fs duration falls off to 7.18$\mu_B$ at 93~fs. At the opposite temporal extreme, the sub-cycle limit (a single cycle of 2.4~eV light corresponds to 1.73~fs) reveals a persisting moment enhancement of up to 1.0$\mu_B$ (in the Supplemental we show that even attosecond pulses can achieve this moment increase). Remarkably, therefore, a robust light induced moment increase is found for over two orders of magnitude of pulse duration. Underlying the less efficient moment enhancement in the sub-cycle regime is the broadband range of frequencies contained in such short pulse envelopes, resulting in excitation also to the Cr $t_{2g}$ minority states, Fig.~\ref{fig3}(c,d). In the long time limit, in contrast, excitation occurs only to the majority Cr $e_g$ states, Fig.~\ref{fig3}(e,f).

Tuning of the pulse central frequency, Fig~\ref{fig3}(b), reveals two distinct  windows: below $\sim 3.2$~eV a light induced increase of moment is observed, while above $\sim 3.2$~eV a decrease in moment is found. This behaviour is found for a broad range of pulse durations and structures, including both linearly and circularly polarized light. The occurrence of moment reduction arises as, at high pulse central frequency, "intraflip" excitations are detuned from the conduction band Cr majority and into conduction band Cr minority states, as may be seen from inspection of the post-pulse density of states, Fig~\ref{fig3}(g-j). This suggests that a sequence of pulses of frequency tuned to the Cr majority and minority conduction bands could effectuate an optical "on-off" switch for moment enhancement, and we find that this is the case, see Supplemental.

\section{{\it Ab-initio} calculation of ultrafast moment increase}

\begin{figure}[t!]
\includegraphics[width=0.95\textwidth]{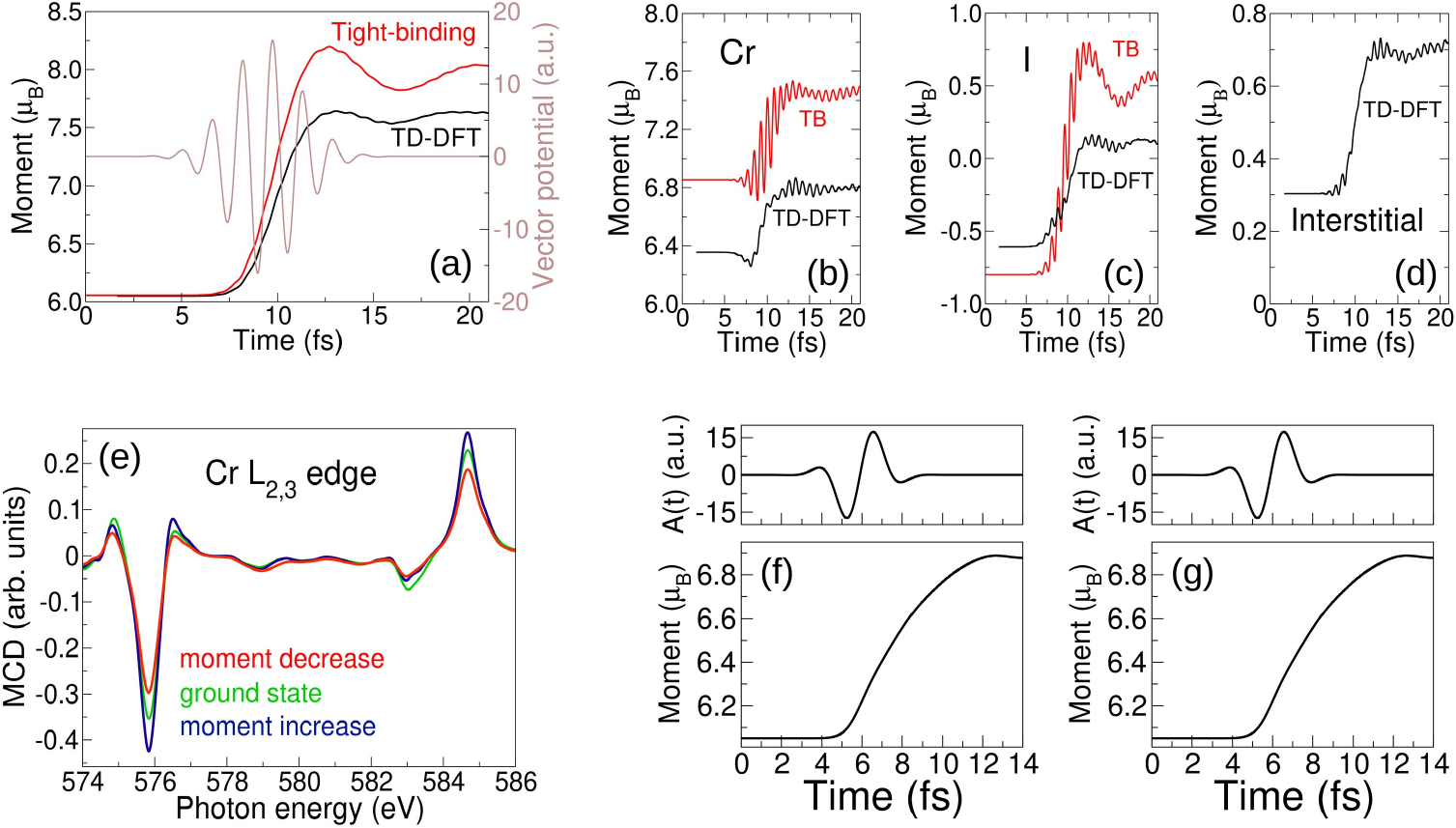}
\caption{\small {\it Time-dependent density functional theory (TD-DFT) calculations of ultrafast moment increase in CrI$_3$}. (a) The temporal dependence of the light induced moment calculated via TD-DFT and the tight-binding method, reveal a similar and substantial ultrafast increase of more than 1.5$\mu_B$. (b-d) The Cr, I, and interstitial components of the moment show qualitatively similar behaviour, with the light pulse inducing a reversal of the I moment. The dichroic response of the Cr L$_{2,3}$ edge, both in the ground state and after ultrafast light induced increase or decrease in moment, is presented in panel (e), with vector potential and transient moment shown in panels (f,g) and (h,i). The clear changes in the dichroic response demonstrate that the Cr L$_{2,3}$ edge provides a sensitive experimental marker of the light induced increase or decrease in moment.
}
\label{fig4}
\end{figure}

We now consider the light induced ultrafast increase in magnetic moment within the very different approach of time dependent density functional theory (TD-DFT). We employ the adiabatic local density approximation and treat the electronic structure within the full potential method as implemented within the Elk code\cite{Elk}; full numerical details are provided in the Supplemental. In contrast to the tight-binding method in which the band structure and spin-orbit coupling remain fixed with dynamical evolution only via changing occupation numbers, full potential TD-DFT allows dynamics of both the effective Kohn-Sham potential -- and thus the spin-orbit coupling -- and the full density $\rho(\v r,t)$. The method thus provides a rigorous test of results obtained via tight-binding calculations.

In Fig.~\ref{fig4}(a) we present the transient moment for a 2.4~eV pulse (vector potential shown in grey) calculated both via the TD-DFT and tight-binding methods. A similar large ($>1.5\mu_B$) increase is seen, with the tight-binding method slightly over estimating the moment increase. A character analysis of the transient moment, Fig.~\ref{fig4}(b-d), reveals a qualitatively similar behaviour in both methods, with a substantial component of the ultrafast moment increase resulting from spin reversal on the I atoms.
Within TD-DFT the magnetization is treated as an unconstrained vector field, which can then be integrated around atomic centres to obtain the species resolved moments. The time dependence of the magnetic moment of the interstitial region, i.e. that part of space not included in the "muffin-tins" around each atomic site, is presented in Fig.~\ref{fig4}(d). The definition of species resolved moments is thus seen to be to some extent arbitrary, accounting for the excellent agreement between the Elk code (full potential basis) and the Quantum Espresso (pseudopotential basis) for the total moment, while differing on species projected moments.

As the TD-DFT calculations employ a full-potential method all states are included in the basis, and this allows for a determination of the x-ray magnetic circular dichroism (XMCD) at the Cr L-edge, which is $\sim580$~eV below the Fermi energy. Transient XMCD spectroscopy provides a sensitive tool of ultrafast changes in moment\cite{bergeard_ultrafast_2014,
willems_optical_2020,
hennecke_angular_2019,
thole_x-ray_1992,
boeglin_distinguishing_2010,
stamm_femtosecond_2007}
, with TD-DFT shown to provide excellent agreement with experiment at early femtosecond times\cite{willems_optical_2020,
dewhurst_element_2020}. The measured ground state XMCD response\cite{frisk_magnetic_2018} can be employed to scissors correct the peak positions, allowing prediction of the transient XMCD as an experimental test. To this end we employ two circularly polarized pulses, that cover the two cases of light induced moment increase and light induced moment decrease, and calculate the XMCD signal after the pulse, Fig.~\ref{fig3}(e). This reveals distinctive changes of the XMCD response to the light induced increase or decrease in moment, thus providing an experimentally testable prediction of the novel moment increase in CrI$_3$. The vector potentials and time dependent moment for the two cases of moment increase and decrease are shown in panels (f,g) and (h,i) respectively.

\section{Ultrafast increase of magnetic moment in CrBrS}

\begin{figure}[t!]
\begin{center}
\includegraphics[width=0.99\textwidth]{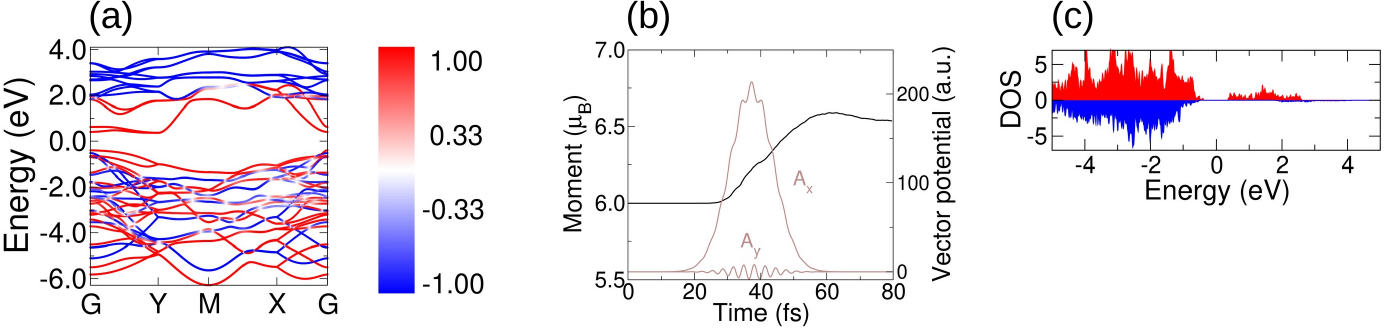}
\caption{\small {\it Ultrafast light-induced moment increase in CrBrS}. (a) The band structure of CrSBr calculated in the Wannierized tight-binding method with band colour indicating the magnitude of the $z$-component of spin (the $x$ and $y$ components are zero). (b)
Action of a light pulse (vector potential shown in grey) generates a moment increase similar to that seen in CrI$_3$, with in panel (c) presented the TD-DOS after excitation by light (red and blue correspond to up and down spin channels respectively).
}
\label{fig5}
\end{center}
\end{figure}

Having established the robustness of our result for CrI$_3$ across distinct methodologies, we now consider the quite different material of CrSBr. Two simple "rules of band design" can be formulated for the "intraflip" mechanism of moment increase to hold: (i) a spin hybridized valence band -- necessary to induce the intra-band spin rotation -- and (ii) spin split and spin pure conduction bands allowing the light pulse to be tuned to allow only spin up or spin down transitions.

CrSBr possesses both requirements (i) and (ii) above, as may be seen from the band structure presented in Fig.~\ref{fig5}(a). The conduction band splitting between spin channels is, however, significantly reduced with the conduction majority and minority spin channels now overlapping. For this material we therefore employ a dual frequency (double pumped) pulse with a large amplitude linearly polarized THz component, designed to induce large intra-band evolution of momentum, and a circularly polarized infra-red component that induces inter-band transitions, pulse vector potential shown in grey in panel (b). This pulse generates a moment increase of 0.5$\mu_B$, falling to 0.4$\mu_B$ 
post pulse, also panel (b), with the transient density of states after the pulse shown in panel (c). This increase signals again the dominance of "intraflips" in the early time spin dynamics when the pulse energy is tuned from a valence spin hybridized band to a spin pure majority conduction band, highlighting the generality of the "intraflip" mechanism of light induced moment increase.

\section{Discussion}

The fact that light couples directly to the momentum of the electron, but not to its spin, naturally invites the picture of a separation of time scales between these processes. Spin preserving charge excitation occurs effectively instantly with a strong laser pulse, in this picture, with spin-orbit induced spin flips occurring after some time delay. Our work presents a strikingly different picture: spin-orbit coupling generates a ground state spin texture in which spin direction is dependent on crystal momentum, implying that the direct light pulse coupling to momentum is also a direct light pulse coupling to spin direction.

This forms the basis of an "intraflip" in which intra-band momentum evolution and inter-band transition combine to excite a valence band state $\ket{\v k\uparrow v}$ to a conduction band state $\ket{\v k\downarrow c}$. This process, as it is driven directly by the light pulse, occurs on a time scale dictated only by the pulse duration, and so can be activated on time scales from sub-femtosecond to hundreds of femtoseconds. Employing tight-binding and state-of-the-art first principles calculations of spin dynamics we have demonstrated lihght induced moment enhancement in the two dimensional magnets CrI$_3$ (2$\mu_B$) and CrSBr (0.5$\mu_B$). The electronic structure required for the effect to occur is, however, general: a spin hybridized valence band and a spin pure up conduction band, implying that this effect will be found in many two- and three-dimensional semi-conducting/insulating magnetic materials. 
While attosecond pulses have theoretically uncovered a moment increase of < 0.1$\mu_B$\cite{neufeld_attosecond_2023} and light induced magnetic order supports a moment increase of < 0.2$\mu_B$\cite{jauk_light_2024}, the lightwave control, magnitude, and robustness, of the magnetic enhancement found here is unprecedented.
Our work thus opens up new avenues of control over magnetism, in which light pulses may effectuate full control of the magnetic moment, on time scales from a single to many optical cycles.

\begin{acknowledgement}

Sharma would like to thank DFG for funding through project-ID 328545488 TRR227 (projects A04), and Shallcross would like to thank DFG for funding through project-ID 522036409 SH 498/7-1. Sharma and Shallcross would like to thank the Leibniz Professorin Program (SAW P118/2021). The authors acknowledge the North-German Supercomputing Alliance (HLRN) for providing HPC resources that have contributed to the research results reported in this paper.

\end{acknowledgement}

\begin{suppinfo}

\subsection{Tight-binding scheme}

The general two-centre tight-binding Hamiltonian is given by

\begin{equation}
H_0 = \sum_{ij} t_{ij} c^\dagger_j c_i
\end{equation}
where $t_{ij}$ is the hopping amplitude between sites $i$ and $j$ (we suppress all other atomic indices). The $t_{ij}$ are obtained via Wannierization using the Quantum Espresso package in conjunction with the Wannier90 software suite, both for CrI$_3$ and CrSBr.

\begin{figure}[t!]
\begin{center}
\includegraphics[width=0.55\textwidth]{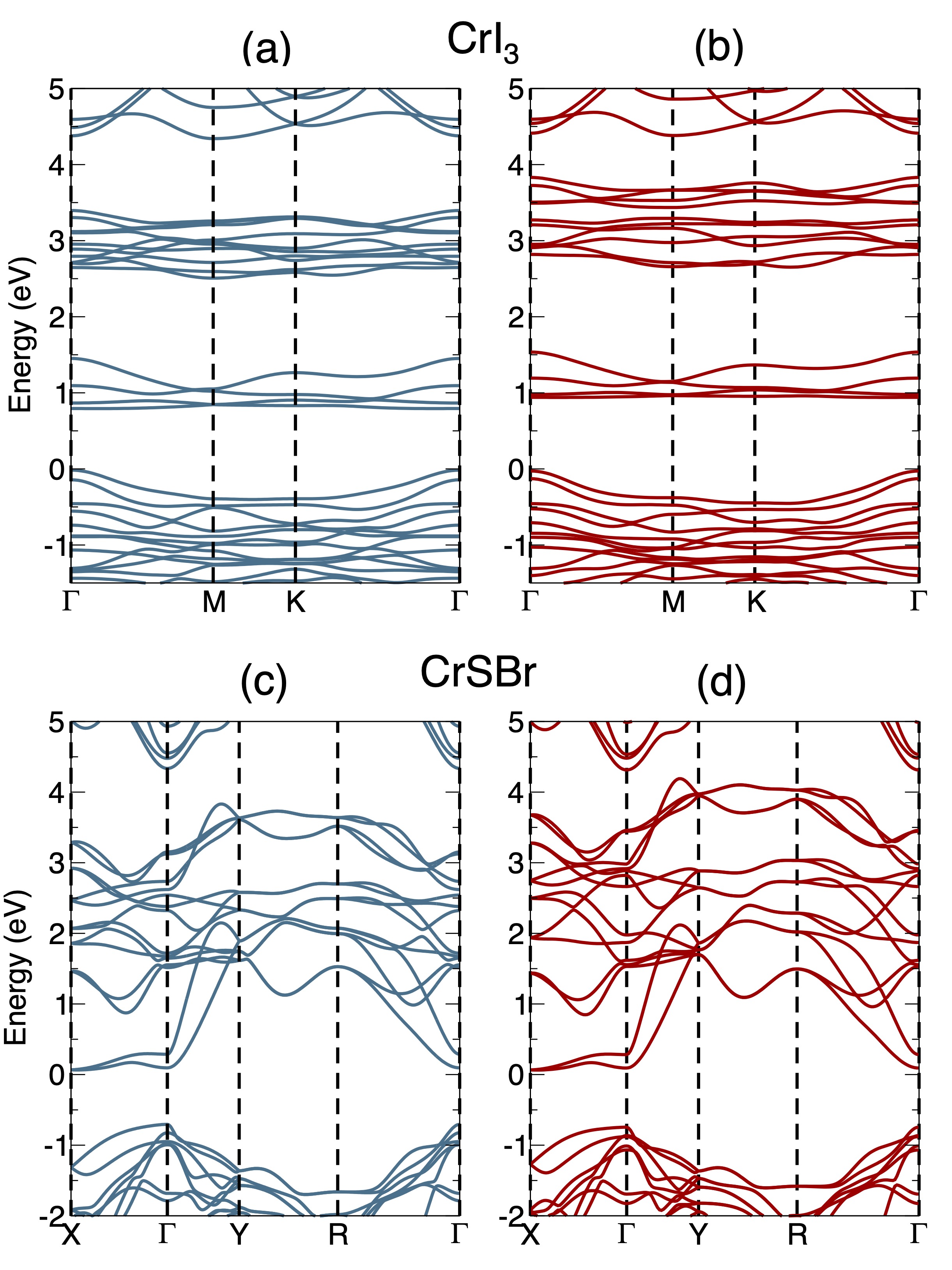}
\caption{{\it Band structures of CrI$_3$ and CrSBr}. Comparison of band structures obtained in the Elk and Quantum Espresso codes for CrI$_3$  (a,b) and CrSBr (c,d). The left hand panel is the band structure obtained via the Elk code, with the right hand panel that obtained via the Quantum Espresso code.
}
\label{S1}
\end{center}
\end{figure}

{\it Numerical parameters for CrI$_3$}: We use a $10\times10\times1$ k-mesh, a vacuum of 20~\si{\angstrom}, and the LDA+U scheme for exchange correlation; $U=4.08$~eV and $J=0.0$~eV are employed which we find reproduces band structures obtained in the literature for this material. The unit cell parameters are $a=b=7.00$~\si{\angstrom} and $c=21.40$~\si{\angstrom}. We Wannierize the $d$ bands of Cr and the $p$ bands of I, yielding a 56 band Hamiltonian.

{\it Numerical parameters for CrSBr}: We use a $15\times15\times1$ k-mesh, a vacuum of 20~\si{\angstrom}, and the LDA scheme for exchange correlation; we find reproduces band structures obtained in the literature for this material. The unit cell parameters are $a=3.54$~\si{\angstrom} and $b=4.75$~\si{\angstrom}, $c=16.42$~\si{\angstrom}. We Wannierize the $d$ bands of Cr and the $p$ bands of S and Br, resulting in 44 bands in the Wannierization.

{\it Tight-binding dynamics}: This time dependent system ket can be expanded in a basis of Wannier states at crystal momentum $\v k(t)$

\begin{equation}
\ket{\Psi_{\v q}(t)} = \sum_n c_{n \v q}(t) \ket{\Phi_{n\v k(t)}}
\label{eq:S1}
\end{equation}
with $\v k(t) = \v q - \v A(t)/c$ given by the Bloch acceleration theorem (with $\v q$ the crystal momentum at $t=0$).

Dynamical evolution is governed by the time-dependent Schr\"odinger equation

\begin{equation}
i \partial_t c_{\v q}(t) = H(\v k(t)) c_{\v q}(t)
\label{eq:cSE}
\end{equation}
where $\v k(t)$ is given by the Bloch acceleration theorem.

\subsection{Time dependent density functional theory}

Real-time TD-DFT \cite{runge1984,sharma2014} rigorously maps the computationally intractable problem of interacting electrons to a Kohn-Sham system of non-interacting electrons in an effective potential. 

Time dependent density functional theory (TDDFT) is an \emph{ab-initio} method for solving the dynamics of many-electron systems via a computationally tractable non-interacting system, known as the Kohn-Sham (KS) system. The non-interacting time-dependent Kohn-Sham (TDKS) equation for periodic systems reads:

\begin{eqnarray}
\label{eq:fullksham}
i\frac{\partial \phi_{j\bk}(\br,t)}{\partial t}=\left[
\frac{1}{2}\left(-i{\boldsymbol \nabla} +\frac{1}{c}{\bf A}\ext(t)\right)^2 \right. 
+v\s(\br,t) + \frac{1}{2c} {\boldsymbol \sigma}\cdot{\bf B}\s(\br,t)  \nonumber \\
+\left.
\frac{1}{4c^2} {\boldsymbol \sigma}\cdot ({\boldsymbol \nabla}v\s(\br,t) \times -i{\boldsymbol \nabla})\right]\phi_{j\bk}(\br,t)
\end{eqnarray}
where $\phi_{j\bk}(\br,t)$ are two-component Pauli spinor TDKS orbitals with quasi-momentum $\bk$, ${\bf A}\ext(t)$ is the external laser field, written as a purely time-dependent vector potential, $\boldsymbol \sigma$ are the Pauli matrices, $v\s(\br,t)=v\ext(\br)+v\H(\br,t)+v\xc(\br,t)$ is the KS effective scalar potential, and ${\bf B}\s(\br,t)={\bf B}\ext(\br,t)+{\bf B}\xc(\br,t)$ is the KS effective magnetic field. The external scalar potential, $v\ext(\br)$, includes the electron-nuclei interaction, while ${\bf B}\ext(\br,t)$ is a external magnetic field which interacts with the electronic spins via the Zeeman interaction.  The Hartree potential, $v\H(\br,t)$ is the classical electrostatic interaction. Finally we have the XC potentials, the scalar $v\xc(\br,t)$, and the XC magnetic field, ${\bf B}\xc(\br,t)$, which require approximation. In this work we used the adiabatic local density approximation (LDA).

{\it Computational parameters for the TD-DFT calculations}: In our calculations of CrI$_3$ we employ a $10\times 10 \times 1$ k-mesh, 75 empty states corresponding to a energy cutoff of 70~eV, and the adiabatic local density approximation (LDA) as our exchange correlation functional $v_{xc}$ and the time step is 2.4 attoseconds. The electronic temperature is set to 300~K. The unit cell dimensions are $a=b=6.89$~\si{\angstrom} and $c=20.0$~\si{\angstrom}.

\cleardoublepage
\pagebreak

\subsection{One dimensional model of moment increase and decrease}

\begin{figure}
\begin{center}
\includegraphics[width=0.55\textwidth]{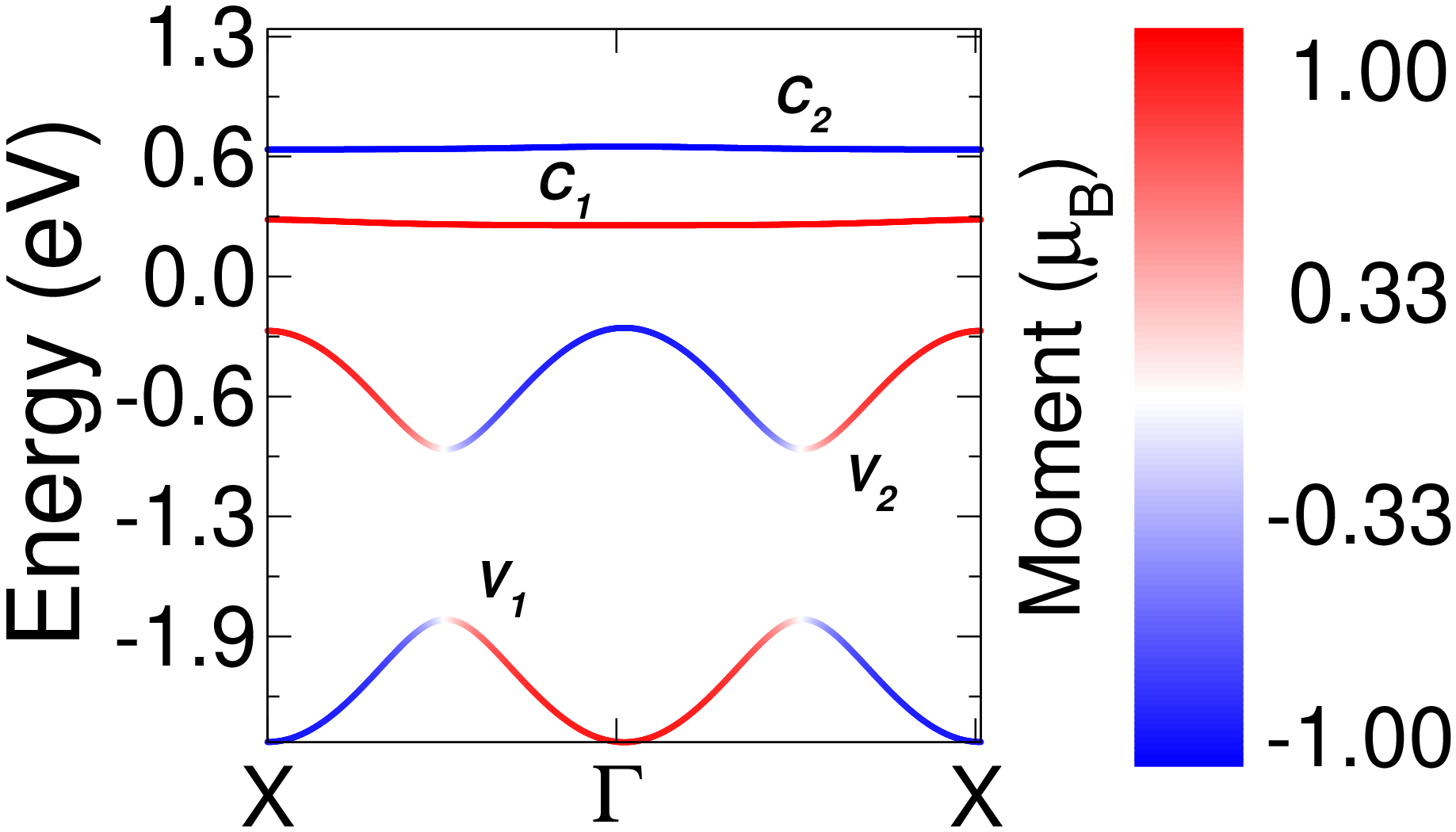}
\caption{{\it Model one dimensional band structure employed to illustrate the "intraflip" mechanism}. The four band model consists of two strolabelngly spin hybridized valence bands, labelled $v_{1,2}$, and two nearly pure spin conduction bands, $c_{1,2}$, with the Fermi level set to zero.
}
\label{S2}
\end{center}
\end{figure}

We consider a four band one dimensional model Hamiltonian consisting of two strongly spin hybridized bands and two nearly spin pure conduction bands, shown in Fig.~\ref{S2}. Here  the spin moment of each eigenstate is shown by the colour as indicated. The following Hamiltonian generates this band structure:

\begin{equation}
H = \begin{pmatrix}
-\alpha_1(\alpha_2 + \cos(a k)) & \beta & \gamma & 0 \\
\beta & \alpha_1(\alpha_2 + \cos(a k)) & 0 & \gamma \\
\gamma & 0 & \alpha_3 + \alpha_5\cos(a k) & 0 \\
0 & \gamma & 0 & \alpha_4 + \alpha_5\cos(a k)
\end{pmatrix}
\label{modham}
\end{equation}
Here the order of the columns are $v_1$ (spin down valence), $v_2$ (spin up valence), $c_1$ (spin up conduction), and $c_2$ (spin down conduction); the diagonal elements therefore represent the energies of these bands, and the off diagonal elements their coupling with $\beta$ controlling the valence band spin hybridization, and $\gamma$ the coupling of valence to conduction. Parameter values that create the band structure shown in Fig.~\ref{S2} can be found in Table~\ref{tabmod}.

\begin{figure}[t!]
\begin{center}
\includegraphics[width=0.95\textwidth]{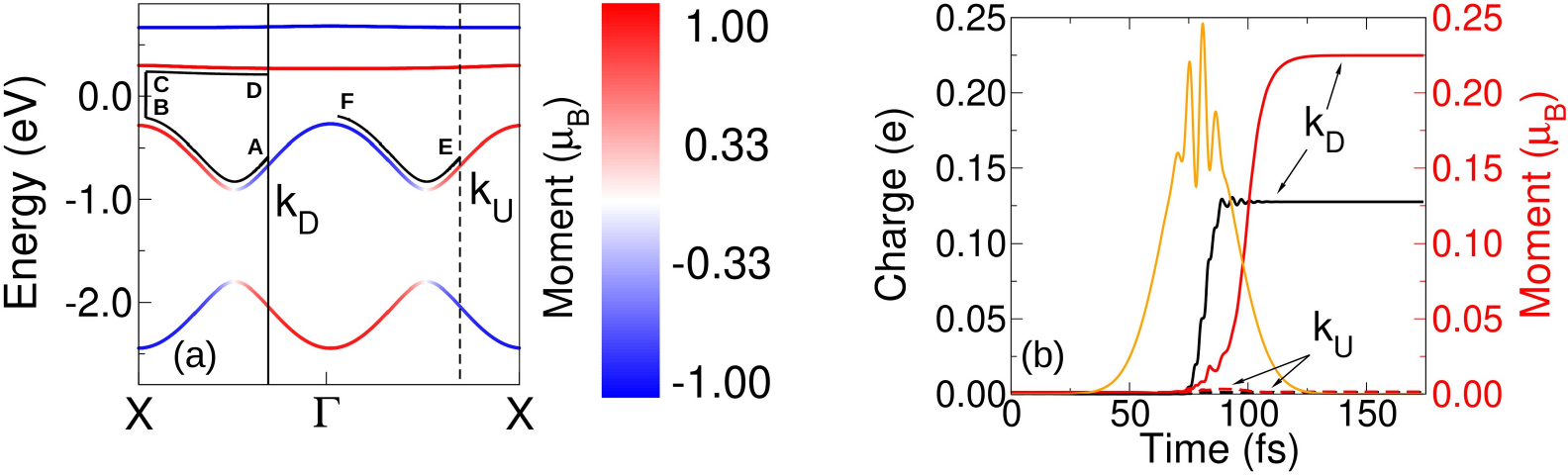}
\caption{{\it Intraflip mechanism of moment control}.
(a) One dimensional model band structure designed to capture the key features of the CrI$_3$ electronic structure: spin hybridized valence bands and nearly spin pure and spin split conduction bands (the Fermi level is set to zero). (b) The charge and moment dynamics at two representative crystal momenta, denoted $k_D$ and $k_U$ in panel (a), with the pulse vector potential driving the dynamics indicated by the orange broken line. The dramatically different spin dynamics of these two momenta -- excitation and significant increase in moment at $k_D$ and almost no change at $k_U$ -- can be understood as the result of the two very distinct dynamical trajectories, indicated by ABCD and EF in panel (a). In the former case intra-band evolution of momenta rotates an initial spin down to spin up at the zone boundary (path AB), allowing optical excitation to the pure up conduction band, with the second pulse half cycle the returning this state to the initial momenta without further rotation, (path CD). The net result is a direct optical excitation A to D that has flipped the spin -- an "intraflip transition". At $k_U$, by contrast, the intra-band evolution of momenta is accompanied by spin rotation from up to down (path EF), precluding direction excitation to the up conduction band, and resulting thus in almost no change from the ground state.
}
\label{S3}
\end{center}
\end{figure}

This model captures the two key features of  CrI$_3$ studied in the main manuscript: (i) strongly spin hybridized valence bands and (ii) strongly spin split conduction bands that are pure up and pure down. We now consider two dynamical pathways within this band structure, at two representative crystal momenta of the excitation labelled $k_D$ and $k_U$, Fig.~\ref{S3}(a). These pathways can be effectuated by a laser pulse consisting of a linearly polarized component resonant with the spin up conduction band augmented by a linearly polarized sub-gap component, vector potential shown in panel (b), a so-called "hencomb" pulse.

We first consider the crystal momenta $k_D$, focusing on the relevant valence band closest to the Fermi energy. The dynamical trajectory, i.e. the pathway of the light induced excitation in momentum-energy space, is schematically indicated by ABCD in Fig.~\ref{S3}(a). The leading edge of the pulse -- dominated by the sub-gap component -- generates intra-band evolution of momenta towards the Brillouin zone boundary (path segment AB) in turn evolving the spin from an initial down state at $k_D$ to a spin up state at the zone boundary. The second pulse component, with energy tuned to the band gap, then generates inter-band excitation into the spin up conduction band, an allowed transition as spin is conserved. As this band manifold is pure up, the second half cycle of intra-band motion (path segment CD) does not further change the spin orientation. The net result is thus an optical transition from valence to conduction at $k_D$ that has flipped the spin. The dynamics of the excited charge (full black line) and $m_z$ moment change (full red line) for this crystal momenta, panel (b), thus reveal an increase in both these quantities.

To bring out fully the role of the sub-gap and gap tuned components of the pulse employed in the manuscript to excite an increase in moment in the model, we present in Fig.~\ref{S4}(a,b) the change in magnetization and excited charge as a function of (i) the sub-gap component amplitude $A_{intra}$ and (ii) the frequency of the gap tuned component $\omega_{ex}$. The crucial role of the former in generating spin flips that allow moment increase can be seen by sending this component to zero: the moment enhancement reduces by a factor of $\sim$5. The excited charge, by contrast, remains almost independent of this pulse parameter.

\begin{figure}[t!]
\begin{center}
\includegraphics[width=0.95\textwidth]{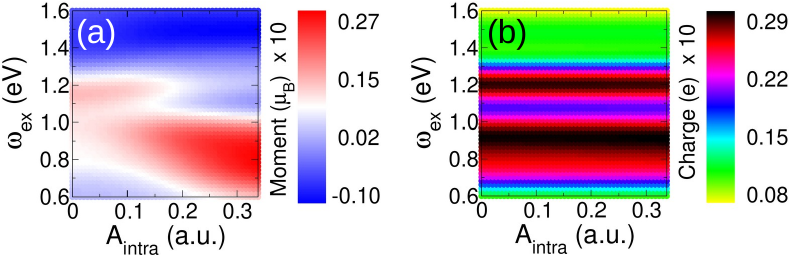}
\caption{{\it Role of pulse components in the intraflip mechanism}. (a) The change in the $z$-component of the spin moment upon light pulse excitation of the band structure shown in Fig.~\ref{S2}. The excitation laser pulse consists of a component whose frequency $\omega_{ex}$ is tuned through the conduction band, and a second sub-gap component of that induces only intraband evolution of momentum with pulse amplitude $A_{intra}$. The change in moment, plotted as a function of $A_{intra}$ and $\omega_{ex}$ reveals (i) significant strengthening of the moment increase as $A_{intra}$ is increased and (ii) a moment change that evolves from moment increase when $\omega_{ex}$ is in resonance with the spin up conduction band to moment decrease when $\omega_{ex}$ is in resonance with the spin down conduction band. In dramatic contrast to this spin physics in which both $\omega_{ex}$ and $A_{intra}$ determine the moment change, the excited charge, panel (b), depends only on the excitation pulse component frequency, $\omega_{ex}$.
}
\label{S4}
\end{center}
\end{figure}

\begin{table}
\begin{tabular}{l|l|l}
\hline
Parameter & Purpose & Value \\ \hline
$\alpha_1$ & Valence band width & 1.00~eV \\
$\alpha_2$ & Valence band separation & 0.00~eV \\
$\alpha_3$ & $C_1$ conduction band centre & 1.60~eV \\
$\alpha_4$ & $C_2$ conduction band centre & 2.00~eV \\
$\alpha_5$ & Conduction band width & 0.00~eV \\
$\gamma$ & Valence-conduction coupling & 0.15~eV \\
$\beta$ & Coupling of valence bands & 0.45~eV
\end{tabular}
\caption{The parameters of the model Hamiltonian, Eq.~\ref{modham}, employed to demonstrate the intraflip concept. $\alpha_{1,4}$ determine band positions and widths, with $\beta$ and $\gamma$ the strength of, respectively, the coupling of the spin up and spin down valence bands $v_1$ and $v_2$, and the coupling of these valence bands to the conduction band system.}
\label{tabmod}
\end{table}

\cleardoublepage
\pagebreak

\subsection{Dependence of moment increase on pulse parameters}

Here we consider the role of the amplitudes of the two pulse components of the "hencomb" pulse employed in the manuscript to CrSBr; the amplitude of the circularly polarized component $A_{circ}$ and the moment of the linearly polarized component $A_{intra}$. We will apply this pulse waveform both to  CrI$_3$ and CrSBr. A special case of this pulse waveform is $A_{intra} = 0$ in which the hencomb pulse reduces to a circularly polarized pulse.
Employing the tight-binding method in Fig.~\ref{S5} and \ref{S6} we present the moment as a function of $A_{circ}$ for a series of values of $A_{intra}$, for CrI$_3$ and CrSBr respectively.

\begin{figure}[h!]
\begin{center}
\includegraphics[width=0.55\textwidth]{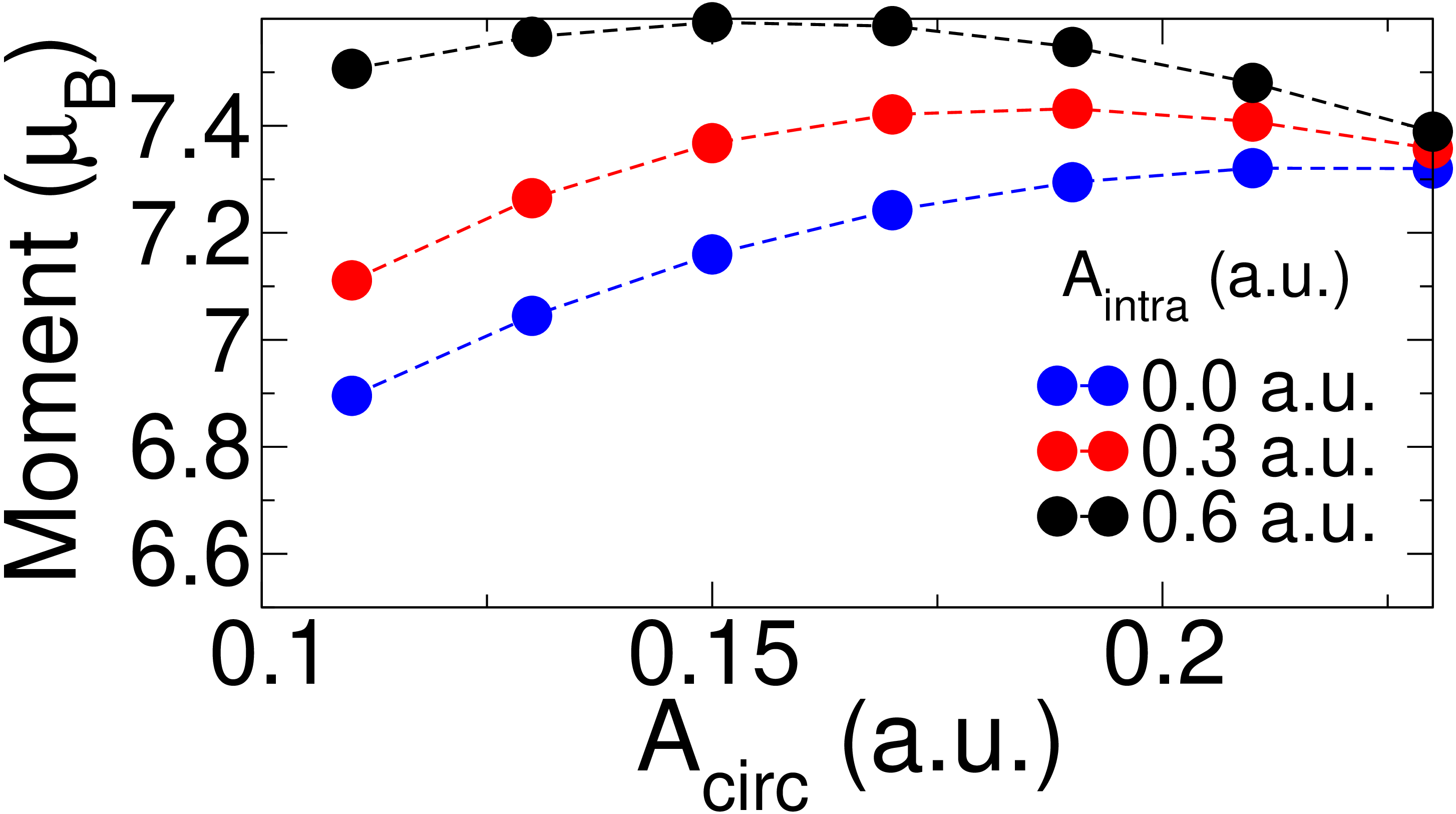}
\caption{{\it The laser excited ultrafast increase in moment of CrI$_3$ presented as a function of pulse component amplitudes; the ground state moment is 6.06$\mu_B$.} We consider the laser pulse presented in Fig.~1(a) of the manuscript, consisting of a circularly polarized component tuned to be in resonance with the valence to spin up conduction band of CrI$_3$ of amplitude $A_{circ}$, and a linearly polarized sub-gap component of amplitude $A_{intra}$, generates a substantial increase of moment over that of the ground state. Increasing $A_{circ}$ enhances the moment increase up to some critical value, after which non-linear processes generate charge excitation to both up and down conduction bands the light induced moment increase weakens. This is shown both for a pure circular pulse, $A_{intra} = 0$, as well as for two "hencomb" pulses of the type employed in the manuscript.
}
\label{S5}
\end{center}
\end{figure}

In Fig.~\ref{S5} it can be seen that as the amplitude of $A_{circ}$ increases the difference in the increase in moment excitation for different values of $A_{intra}$ decreases: the circularly polarized pulse is "taking over" the work of the sub-gap linearly polarized component at large amplitude, generating itself sufficient intraband evolution of crystal momentum to drive the intarflip process. As can be seen, at the high amplitude end of $A_{circ}$ an increase of more than 1~$\mu_B$ can be achieved by circularly polarized light acting alone.

A similar effect can be seen in CrSBr, in which a low $A_{circ}$ the impact of the linearly polarized pulse is dramatic, but this reduces on increase of $A_{circ}$. However, driving the amplitude of the circularly polarized pulse to the high fluence limit results in excitation of charge to both conduction up and conduction down, with eventually the latter resulting in a decrease in moment from the ground state value of 6~$\mu_B$.

\begin{figure}[t!]
\begin{center}
\includegraphics[width=0.5\textwidth]{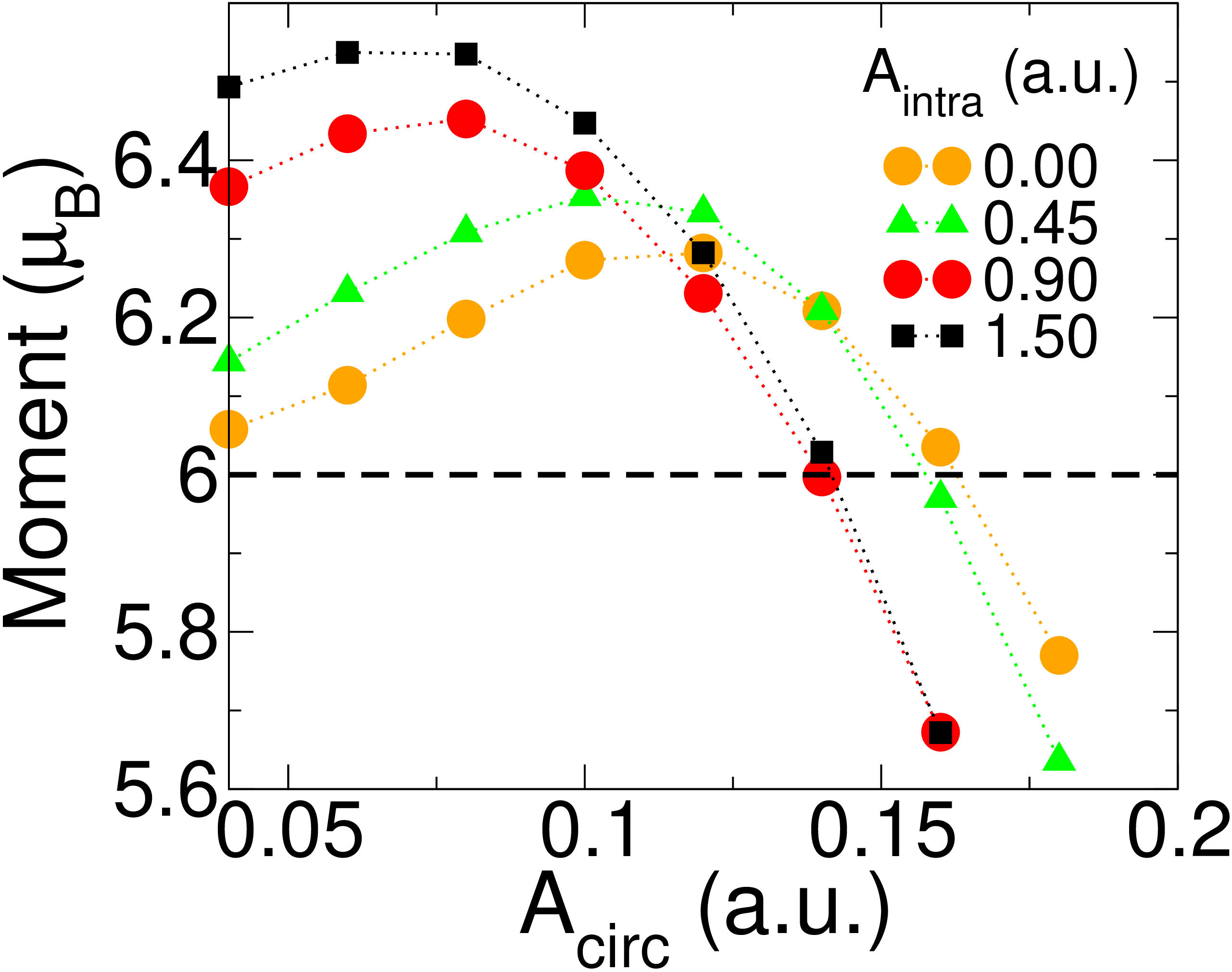}
\caption{{\it The laser excited ultrafast increase in moment of CrSBr presented as a function of pulse component amplitudes; the ground state moment is 6.0$\mu_B$, indicated by the broken line.} As for CrI$_3$, Fig.~\ref{S5}, we present the moment variation with the amplitude of the spin up conduction band tuned circularly polarized pulse $A_{circ}$, for a series of values of the amplitude of the sub-gap linearly polarized component, $A_{intra}$.
}
\label{S6}
\end{center}
\end{figure}

\cleardoublepage
\pagebreak

\subsection{Comparison of TD-DFT and tight-binding for linearly polarized pulses}

\begin{figure}[h!]
\begin{center}
\includegraphics[width=0.95\textwidth]{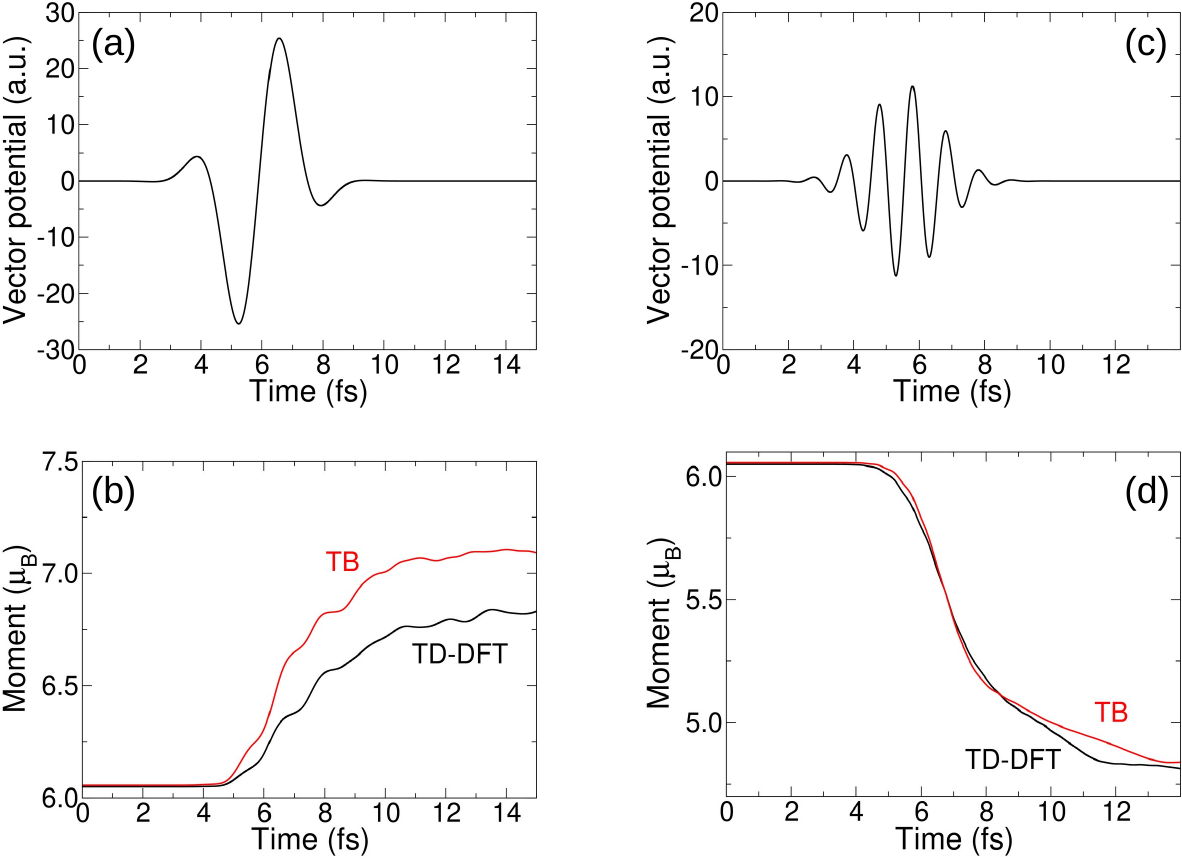}
\caption{{\it Comparison of time dependent density functional theory (TD-DFT) and tight-binding (TB) methods for calculating the transient moment}. Shown are (a) a linearly polarized pulse that generate a moment increase, panel panel (b), and a higher frequency pulse, panel (c), that generates moment decrease, panel (b). The transient moment is seen to be similar qualitatively similar behaviour when calculated by the TB and TD-DFT methods.
}
\label{S7}
\end{center}
\end{figure}

\cleardoublepage
\pagebreak

\subsection{Ultrafast moment control in CrI$_3$}

\begin{figure}[h!]
\begin{center}
\includegraphics[width=0.95\textwidth]{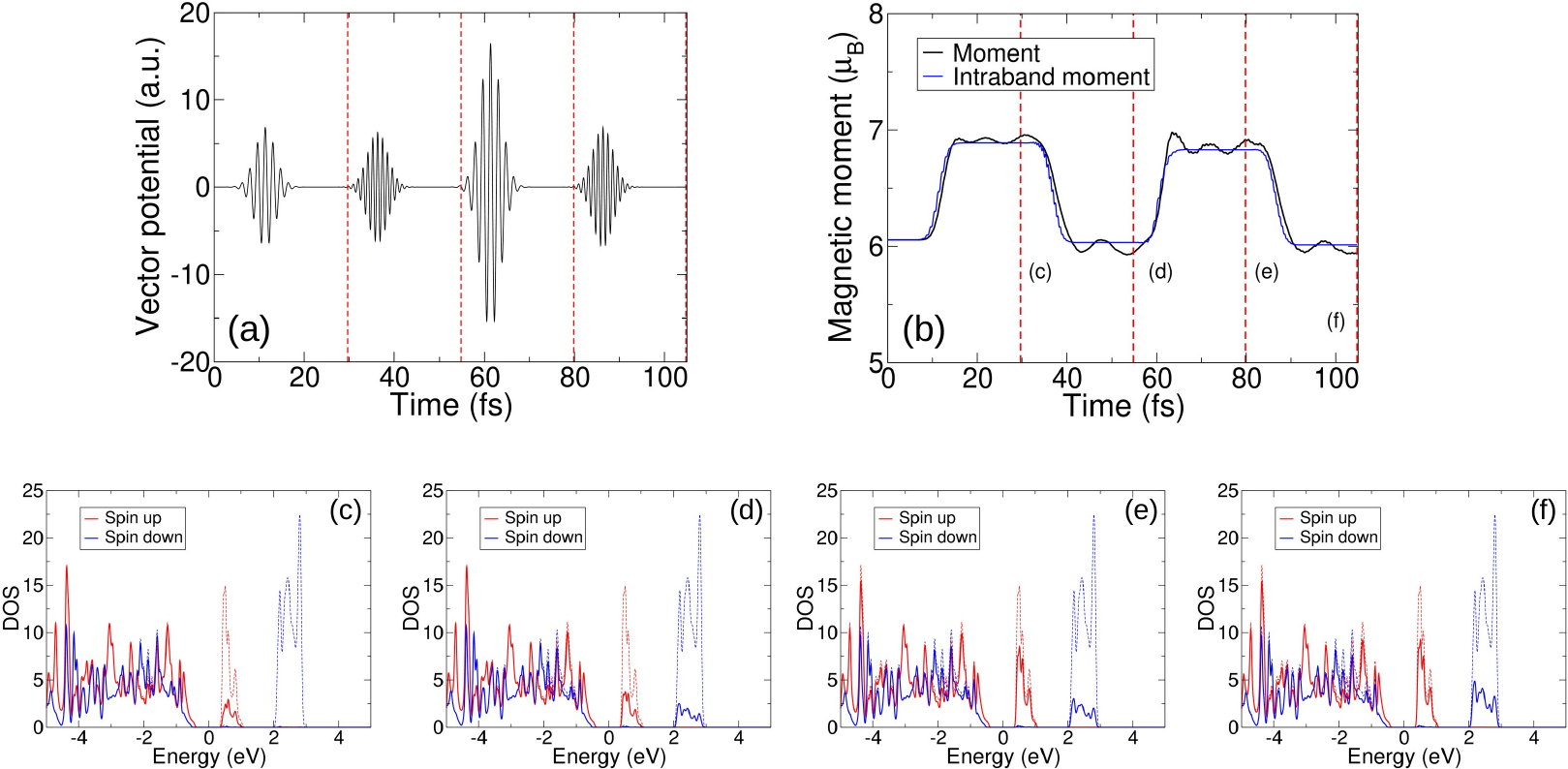}
\caption{{\it Controlling the transient moment in CrI$_3$}. Successive and temporally separated pulse components tuned either to the majority Cr conduction $t_2g$ or minority Cr conduction $e_g$, vector potential shown in panel (a), generate successive increase and decrease of magnetic moment in CrI$_3$, panel (b), via the "intraflip" mechanism discussed in the main paper. At times corresponding to the vertical red lines the transient density of states is shown, panels (c-f) as indicated, revealing the expected successive population of the majority and minority conduction states. The dashed lines indicate the density of states and the full line the occupied density of states.
}
\label{S8}
\end{center}
\end{figure}

\cleardoublepage
\pagebreak

\subsection{3-band model}

\begin{figure}[h!]
\begin{center}
\includegraphics[width=0.95\textwidth]{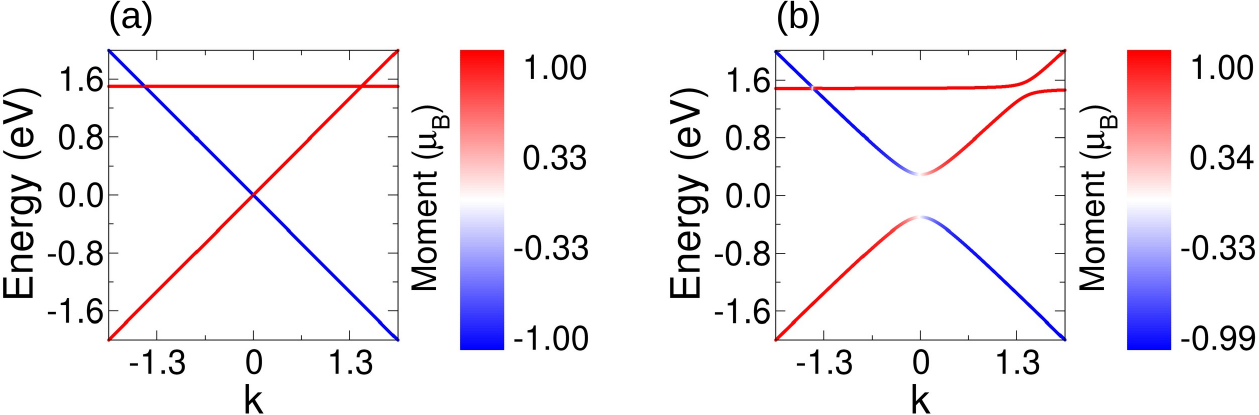}
\caption{{\it Model band structure}. (a) Band structure with $\Delta = \gamma= 0$ (b) Band structure for finite $\Delta$ and $\gamma$.
}
\label{S9}
\end{center}
\end{figure}

Intersecting spin up and down bands that hybridize via spin orbit-coupling generate bands that exhibit a continuous evolution between spin up and spin down states. This is illustrated in Fig.~\ref{S9} via the 3-band Hamiltonian

\begin{equation}
H = \begin{pmatrix}
-\alpha k & \Delta & 0 \\
\Delta & \alpha k & \gamma \\
0 & \gamma & \epsilon
\end{pmatrix}
\end{equation}
where $\pm \alpha k$ represent two intersecting bands of opposite spin, with $\Delta$ a coupling between them. The high energy band at  $\epsilon$ couples to these via the parameter $\gamma$. In panel (a) is shown the case of $\Delta=\gamma=0$, i.e decoupled bands, and panel (b) the case in which these parameters are finite and both spin hybridization and coupling of these hybridized bands to the $\epsilon$ band can be seen.
 
This texture, as described in the main text, supports a three stage pathway that excites from a valence spin down state to conduction spin up state. This consists of: (i) a light driven intra-band evolution of momentum that rotates an initial spin down state to up, followed by (ii) inter-band excitation to the spin up conduction band, with (iii) subsequent intra-band evolution returning the state to its initial momentum while preserving the spin up state.

\begin{figure}[t!]
\includegraphics[width=1.0\textwidth]{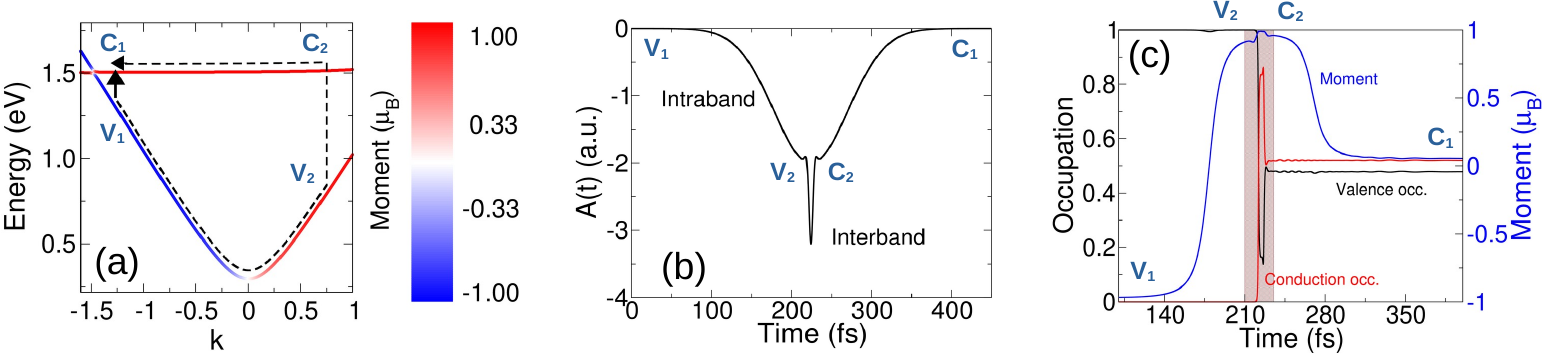}
\caption{\small {\it Intraflip mechanism of moment control}.
(a) A model band structure exhibiting spin hybridization and the evolution between spin up to spin down within both valence and conduction bands. Two possible light induced excitations within this electronic structure are indicated: a direct optical excitation from valence to conduction, $V_1 \to C_1$. This, however, is forbidden as $V_1$ and $C_1$ have opposite spin. A second "intraflip" excitation is indicated by the dashed lines $V_1 \to V_2 \to C_2 \to C_1$. This excitation involves first an intra-band evolution from spin down to spin up ($V_1 \to V_2$) followed by an allowed spin up to spin up transition ($V_2 \to C_2$). After intra-band evolution $C_2 \to C_1$ this results in a spin flip excitation $V_1 \to C_1$, mediated by intra-band evolution of crystal momentum. (b) A model laser pulse that generates this "intraflip" excitation consisting of two Gaussian envelopes. The weak electric field component, labelled "intraband", drives the required intra-band evolution of crystal momentum, but does not generate inter-band excitation. A second component possessing a strong electric field drives inter-band excitation from valence to conduction, labelled "interband". (c) The change in occupation and moment during this "intraflip" process. Between $V_1$ and $V_2$ there is no change in occupation numbers -- the valence band occupation is 1 and the conduction 0 -- while the spin moment rotates from down to up. This is followed by excitation to the conduction band, indicated by the shaded area, in which the conduction occupation rises to $\sim 0.5$. Finally, in the second half cycle of the Gaussian pulse the moment reduces as the valence band state rotates back from up to down (along $V_2 \to V_1$) while the conduction band moment remains up ($C_2 \to C_1$). The overall effect is therefore a moment increase.
}
\label{S10}
\end{figure}

To demonstrate this mechanism we consider dynamical evolution within this three-band model. In Fig.~\ref{S10}(a) we show a zoom of the band structure of this model with a dynamical pathway indicated. We take the Fermi level to be at 1.4~eV, thus $V_{1,2}$ are valence states and $C_{1,2}$ conduction band states. A direct optical excitation from valence to conduction, $V_1 \to C_1$ as indicated by the full arrow, is forbidden as the states $V_1$ and $C_1$ are spin down and spin up respectively. An "intraflip" excitation pathway, indicated by the dashed line $V_1 \to V_2 \to C_2 \to C_1$, is generated by the pulse shown in Fig.~\ref{S10}(b). This consists of two Gaussian envelopes, the first of which, labelled "intraband", drives an evolution of momentum from $V_1 \to V_2$ and back without generating inter-band transitions while the second, labelled "interband", generates inter-band excitation from $V_2$ to $C_2$. During the initial intra-band excitation valence and conduction occupation are fixed at 1 and 0 respectively, while the spin moment evolves from down to up, panel (c). At exactly half-cycle, the shaded region, excitation from valence to conduction occurs, allowed as both states are spin up, with the conduction and valence occupancy now both $\sim 0.5$. Finally, the second half-cycle returns the crystal momentum to its initial value, with the overall result a spin flip excitation $V_1 \to C_1$.

\cleardoublepage
\pagebreak

\subsection{Attosecond moment enhancement}

\begin{figure}[h!]
\begin{center}
\includegraphics[width=0.95\textwidth]{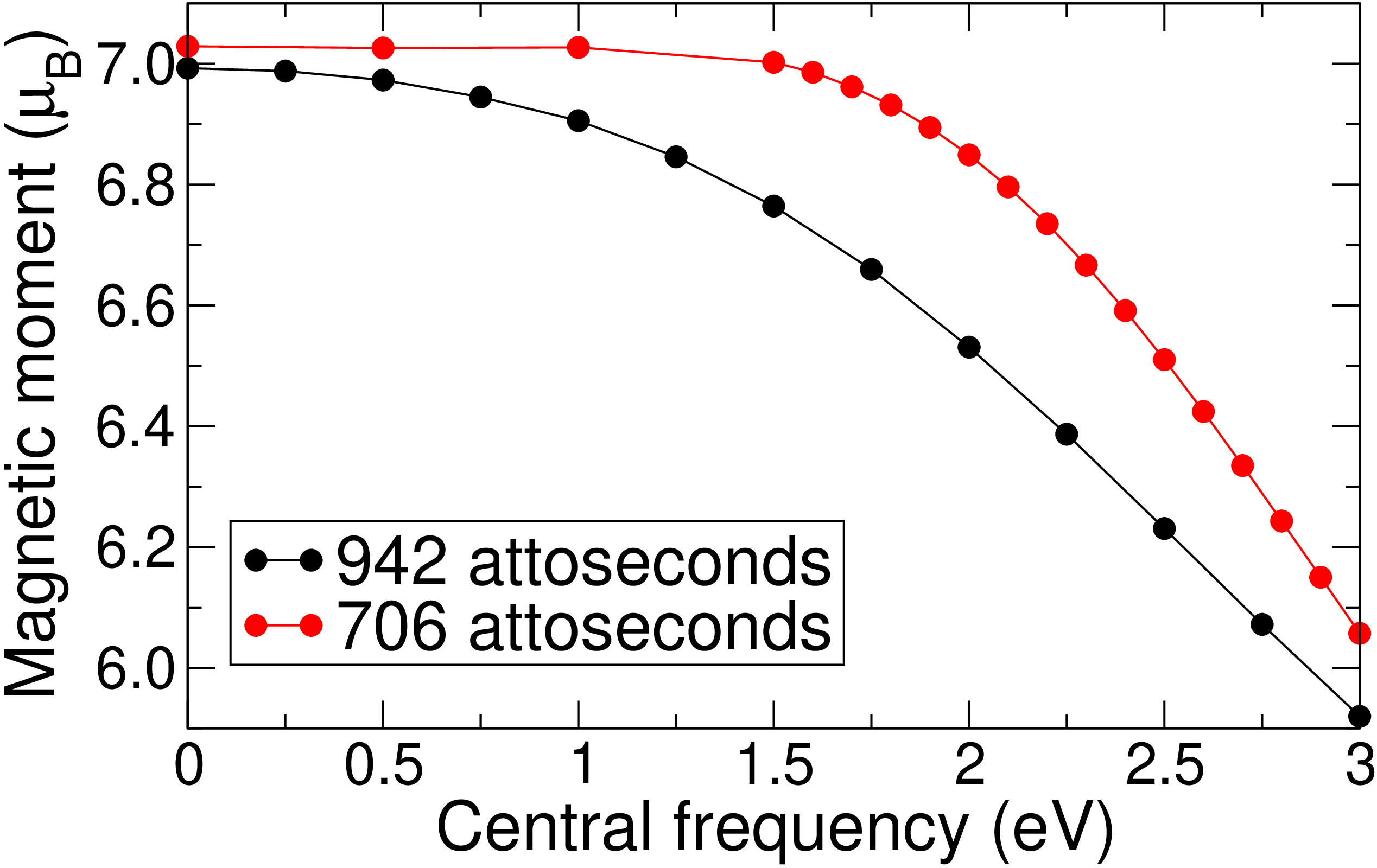}
\caption{{\it Light induced moment enhancement in CrI$_3$ in the attosecond regime}. By reducing the pulse central frequency the moment enhancement is continuously increased as intraflip excitations to the minority Cr conduction band are supressed. As may be seen, for both a 942 and 706 attomsecond pulse duration a substantial moment enhancement of 1~$\mu_B$ is obtained.
}
\label{S11}
\end{center}
\end{figure}

\end{suppinfo}

%\bibliography{current,samItems,2dm,xmcd}

\providecommand{\latin}[1]{#1}
\makeatletter
\providecommand{\doi}
  {\begingroup\let\do\@makeother\dospecials
  \catcode`\{=1 \catcode`\}=2 \doi@aux}
\providecommand{\doi@aux}[1]{\endgroup\texttt{#1}}
\makeatother
\providecommand*\mcitethebibliography{\thebibliography}
\csname @ifundefined\endcsname{endmcitethebibliography}
  {\let\endmcitethebibliography\endthebibliography}{}

\end{document}